\documentstyle[epsfig,longtable]{aipproc}

\begin{document}
\title{Looking for GRB progenitors}

\author{Krzysztof Belczy{\'n}ski$^1$, Tomasz Bulik$^1$ and 
        Bronis{\l}aw Rudak$^{2,3}$}
\address{
$^1$Nicolaus Copernicus Astronomical Center,
Bartycka 18, 00-716 Warszawa,Poland\\
$^2$ Nicolaus Copernicus Astronomical Center,
Rabia{\'n}ska 8, 87-100 Toru{\'n},Poland\\
$^3$ Toru{\'n} Center for Astronomy, Nicolaus Copernicus 
University, Gagarina 11, 87-100 Toru{\'n}, Poland}

\maketitle

\begin{abstract}
Using stellar binary population synthesis code we calculate the
production rates and lifetimes of several types of possible GRB
progenitors. We consider mergers of double neutron stars, black
hole neutron stars, black hole white dwarfs and helium star
mergers.  We calibrate the results with the measured star
formation rate history.   We discuss the viability of each GRB
model, and alternatively assuming that all bursts are connected
with one model we constrain the required collimation of GRBs. We
also show the importance of widely used evolutionary parameters
on  the merger rates of calculated binary populations.
\end{abstract}

\section*{Introduction}

It is surprising how little we know
of GRB progenitors when we consider how much work
is devoted to the subject. Different objects and models of progenitors were
proposed, some already forgotten while others still being intensively studied.  
Considering the diversity of gamma ray bursts it is probable that they come 
from different types of astronomical objects, and this should promote the work
on different types of proposed progenitors. Recently much of the weight was
placed on collapsars, and connection  between supernovae and GRBs, although
other models, as compact object  mergers, are still on the stage. 

Compact object binaries have recently drawn much attention in the astronomical
community. Several models of compact object binaries were proposed as GRBs
progenitors, namely: double neutron star mergers 
\cite{meszrees97,ruffert97}, black hole -- neutron star mergers 
\cite{lee95}, black hole -- white dwarf mergers \cite{fryer99b} and  helium
star mergers \cite{fryer98}. Mergers of compact object binaries are also most
often considered sources of gravitational waves. As gravitational wave
detectors, LIGO and VIRGO, will soon be operational,  the question of these
merger rates arises \cite{allen99}.

To predict compact object merger rates we use Monte Carlo simulations to
produce populations of different compact object binaries.
In this approach one generates massive binaries at ZAMS and evolves them
through consecutive stages of single and binary star evolution, which may
eventually lead to formation of a compact object binary.
Final synthesis of large ensembles of compact object binaries allows then
statistical studies and calculation of expected merger rates.

Several compact object merger
rates estimates \cite{narayan91,phinney91,tutukov93,zwart96b,bb1,bbz,fryer99a}
have been already published. However,
calculations of this rates are based on many assumptions and use parameters
with very uncertain values. Whereas the  evolution of single stars is reasonably well
known \cite{eggleton89}, the distributions of initial binary conditions as well
as some aspects of binary star evolution are uncertain. Finally the population
synthesis codes must deal with uncertainties in supernova explosion mechanisms,
and in particular with the value of the kick a neutron star (or a black hole)
receives at birth, and the mass of formed in the explosion compact object
\cite{bbz}.
Previous population synthesis calculations concerned  double neutron stars and 
only few dealt with neutron star black hole binaries.
So far only in  one case \cite{fryer99a} all kinds of the proposed GRBs
compact object binary progenitors were considered.
Our present contribution follows the same line of work, although we use
a different population synthesis code and we would like to communicate
our first results here. 
Preliminary comparison shows striking similarities of the results, which
taking into account different codes, points toward robustness of the
population synthesis method in spite of  many uncertainties involved in
the calculations.

\section*{The Model}

Our evolutionary code is primarily based  on the prescriptions given by 
\cite{tout97} and \cite{eggleton89}, but we use the number of many
revised or newly developed specific evolutionary prescriptions from 
\cite{bethe98,zwart96a,pols94,podsiadlowski92}.
In our calculation of single and binary evolution we include stellar winds
(normal and LBV wind), magnetic breaking, quasi-dynamic mass transfer,
common envelope evolution, hyper-accretion onto compact objects, detailed
supernova explosion treatment and gravitational wave energy loss in compact
object binaries.
We start the evolution of a given binary when both components are at ZAMS, then
each star evolves through different stages of its life depending on its mass: 
main sequence, Hertzsprung gap, red giant branch, horizontal branch, asymptotic
giant branch, and either supernova explosion which leads to formation of a
neutron star (or a black hole)  or a phase of enhanced mass loss and  formation
of a white dwarf. During each stage of the binary evolution the components may 
interact, which changes the consecutive component's evolution (either through
rejuvenation, stripping the component off its outer layers, or even by
swallowing one component by the other). Interaction of binary components may
lead to one of the proposed GRB progenitors, the helium star merger, when the
compact object is engulfed  in giant's envelope. If there is not enough orbital
energy to eject the  giant's  envelope then the compact object spirals in
through the giant's envelope finally merging with the giant's core, which can
be torn out in the process, and form an  accretion disc around a compact object
which is then already a black hole,  due to the accretion during spiral in.  

Other systems follow their evolutionary paths, and we collect the
informations on the types of compact object binaries that are proposed 
for GRBs progenitors: double neutron stars, black hole neutron stars and 
black hole white dwarf systems.
These systems interact only due to gravitational energy wave loss,
finally  merging and forming a black hole with massive thick
accretion disc which presumably may lead to a gamma ray burst.

\section*{Results}

In Fig. 1 we show the relative numbers of 4 different GRB progenitor types that 
merge within the Hubble time (15 Gyrs) as a function of the width of the
distribution from which we draw kick velocity a compact object receives in a
supernova explosion.
Two things are clearly seen; first the number of WD-BH binaries and Helium
mergers (He-BH) is about the same and is more then an order of magnitude
greater then the number of NS-NS and BH-NS binaries. Second, the number of a
given progenitor type falls off approximately exponentially with the kick 
velocity.
Relative production rates may be calibrated (eg. see eq. 14 in \cite{bb1}).
For example, assuming the width of kick velocity, let say, 
v$_{kick}=200$ km s$^{-1}$ yields:  
1 merging event per Milky Way like galaxy per Myr for BH-NS systems, 
3 events for NS-NS binaries and 60 for WD-BH and Helium mergers.

\begin{figure}[t!] 
\vspace*{-0.5cm}
\centerline{\epsfig{file=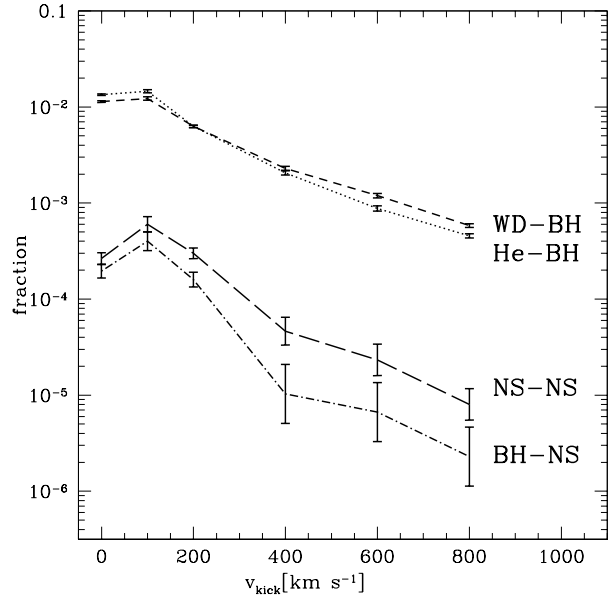,width=3.3in}}
\vspace{10pt}
\caption{Relative production rates 
 of different GRB progenitor types that merge within the Hubble time.}
\label{fig1}
\end{figure}

\begin{figure}[t!] 
\vspace*{-0.5cm}
\centerline{\epsfig{file=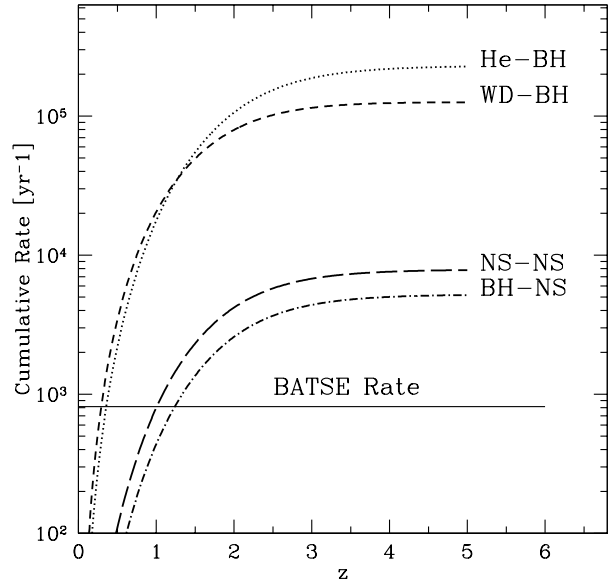,width=3.3in}}
\vspace{10pt}
\caption{Cumulative merger rates for different GRB progenitor 
types as a function of redshift.}
\label{fig2}
\end{figure}

In Fig. 2 we show the cumulative rate of different merging events.
We have combined our relative numbers for different progenitors with the
star formation rate function \cite{madau96,totani97}, and taking into account 
the evolutionary time delay of a given merging event we integrated our
production rates to get the merger rates as a function of redshift.
In this example calculation we assumed 
$\Omega_M=1.0, \,\, \Omega_\Lambda = 0.0 $, and the 
 Hubble constant $H_0 =65$ km s$^{-1} {\rm Mpc}^{-1}$.
We used the kicks drawn from the distribution which is a
weighted sum of two Gaussians: 80 percent with the width of 200 km s$^{-1}$ 
and 20 percent with the width 800 km s$^{-1}$.

In Fig. 2 we show also BATSE gamma ray burst detection rate corrected
for the sky exposure.
Comparison of the cumulative distributions for different progenitor types with
he BATSE rate shows that if any of the progenitor types included here  was to
reproduce the BATSE rate, 
then we should not see GRBs from redshifts greater then unity! 
Of course this is not the case, as  GRBs with higher redshifts
were observed.
However, we haven't yet introduced the collimation factor into our results,
presented in Fig. 2, which certainly will lower down our predicted rates.
To lower down our calculated rates to the BATSE rate, for average GRBs 
redshift of about 2, we would need the collimation of about 4$^\circ$ for 
BH-WD and Helium mergers and about 12$^\circ$ for BH-NS and NS-NS mergers.   
 
As seen in Fig. 2 the curves flatten out for high redshifts (z $\geq$ 5).
In other words we do not expect binary mergers at high redshifts.
This is a combined effect of the star formation rate function we have used, 
which falls down for high redshifts and of the 
non zero lifetimes of progenitors, prior to the final merging event. 
Each of our binaries needs a specific time to evolve into a 
compact object binary (t$_{evol}$) and then needs time to merge due to gravitational wave energy 
losses (t$_{merger}$). 
This times are non negligible and are specific for each group of proposed
binary GRBs progenitors.
For our sample of binaries we found that $t_{life}$\ ($t_{evol}+t_{merger}$) are,
for NS-NS: $\sim 10^7$--$10^{12}$\ yrs,
for BH-NS: $\sim 10^7$--$10^{10}$\ yrs, 
for BH-He: $\sim 10^6$--$10^9$\ yrs, 
for BH-WD: $\sim 10^7$--$10^{12}$\ yrs.

\section*{Conclusions}

\noindent
\begin{itemize}
\item GRB binary progenitor production rates fall off exponentially
with width of natal kick velocity distribution.
\item Evolutionary times can not be neglected in computations of
GRB's progenitor rates.
\item Assuming that all GRB's are connected to NS-NS or BH-NS
binaries, the collimation must be of order $\sim 3 \times 10^{-2}$
(12$^\circ$). If we 
assume that all GRB's result from binary mergers, then the
population is dominated by BH-WD and BH-He star mergers,
and the collimation must be of order $\sim 3 \times 10^{-3}$
(4$^\circ$).
\item We do not expect binary mergers at high redshifts (z $\geq$ 5).
\end{itemize}

{\bf Acknowledgments.} We acknowledge the support of the following
grants KBN-2P03D01616, KBN-2P03D00415, KBN2P03D02117.

\end{document}